\documentclass[prb,twocolumn,showpacs,preprintnumbers,amsmath,amssymb]{revtex4}
\usepackage{graphicx}
\usepackage{longtable}
\usepackage{dcolumn}
\usepackage{bm}

\begin{document}

\title{Strong magnon softening in tetragonal FeCo compounds}
\author{Ersoy \c{S}a\c{s}{\i}o\u{g}lu}\email{e.sasioglu@fz-juelich.de}
\author{Christoph Friedrich}
\author{Stefan Bl\"{u}gel}

\affiliation{Peter Gr\"{u}nberg Institut and Institute for
Advanced Simulation, Forschungszentrum J\"{u}lich and JARA, 52425
J\"{u}lich, Germany}

\date{\today}

\begin{abstract}

Magnons play an important role in fast precessional magnetization
reversal processes serving as a heat bath for dissipation of the
Zeeman energy and thus being responsible for the relaxation of
magnetization. Employing \emph{ab initio} many-body perturbation
theory we studied the magnon spectra of the tetragonal FeCo
compounds considering three different experimental $c/a$ ratios,
$c/a=$1.13, 1.18, and 1.24 corresponding to FeCo grown on Pd, Ir,
and Rh, respectively. We find that for all three cases the
short-wave-length magnons are strongly damped and tetragonal
distortion gives rise to a significant magnon softening. The
magnon stiffness constant $D$ decreases almost by a factor of two
from FeCo/Pd to FeCo/Rh. The combination of soft magnons together
with the giant magnetic anisotropy energy suggests FeCo/Rh to
be a promising material for perpendicular magnetic recording
applications.
\end{abstract}


\maketitle

Since the introduction of the first commercial hard disk drive
in 1956, the recording density in a hard disk , that is the amount
of information that can be stored per square inch, has increased
by more than 7 orders of magnitude to meet an ever-growing
need.\cite{IBM} This has been achieved by a simple scaling of the
dimensions of the bits recorded in storage medium. Due to the
superparamagnetic effect, however, the recording density has an
upper limit. For longitudinal magnetic recording it is around 200
gigabit per square inch, whereas it is predicted to be much larger
for perpendicular recording, up to 1000 gigabit per square inch,
though this limit is constantly changing with the discovery of new
materials.\cite{Fontana,Piramanayagam,Plos}

The major problem in designing magnetic storage media is to retain
the magnetization of the medium over a long period of time despite
thermal fluctuations. If the ratio of the thermal energy
$k_\mathrm{B}T$ to the magnetic energy per grain $K_\mathrm{u}V$,
where $V$ is the grain volume and $K_\mathrm{u}$ is the uniaxial
magneto-crystalline anisotropy energy, becomes sufficiently large,
the thermal fluctuations can reverse the magnetization in a region
of the medium destroying the data stored
there.\cite{Piramanayagam,Weller} In order to further increase the
recording density in future recording media, high-$K_\mathrm{u}$
materials are needed.\cite{Weller2} Additionally, a large
saturation magnetization $M_\mathrm{s}$ is beneficial to reduce
the write field, which has to be applied by the writing head.
Materials that combine the desired large values of $K_\mathrm{u}$
and $M_\mathrm{s}$ are tetragonal near-equiatomic FeCo alloys. The
large values of $K_\mathrm{u}$ and $M_\mathrm{s}$ in these alloys
were first predicted by first-principles
calculations\cite{Burkert} and then confirmed by
experiments.\cite{Andersson,Luo,yildiz_1,yildiz_2} In particular,
Yildiz \textit{et al.}\cite{yildiz_2}  achieved a strong
perpendicular magnetic anisotropy (PMA) in tetragonal FeCo alloys
epitaxially grown on Pd ($c/a=1.13$), Ir ($c/a=1.18$), and Rh
($c/a=1.24$) substrates. The authors found that the PMA is very
sensitive to the tetragonal distortion and increases with
increasing $c/a$ ratio, which allows to tune the PMA by growing
the alloys on different substrates.

Besides large $K_\mathrm{u}$ and $M_\mathrm{s}$ values, another
very important issue in magnetic recording applications is the
magnetic switching time, which imposes physical limits on data
rates and areal recording densities.\cite{Plumer} In current
devices the switching speeds have reached a point where dynamical
effects are becoming very
important\cite{Plumer,Visscher,Safanov,Fidler,Garanin} Magnons are
created in fast (field driven) as well as ultrafast (laser
induced) magnetization reversal
processes.\cite{Silva,Feygenson,Lenz,Baberschke,Zakeri,Munzenberg_1,Munzenberg_2,Rasing,Koopmans,Manchon}
The former case is of particular interest for current device
applications. It is found that above some threshold magnetic field
the uniform precessional mode, i.e., the $\mathbf{k}=0$ magnons
decay into nonuniform magnons ($\mathbf{k} \neq 0 $), i.e., the
Zeeman energy stays in the magnetic subsystem and scatters between
magnon modes.\cite{Silva,Feygenson,Lenz,Baberschke} However, in
ultrafast magnetization reversal the high-energy electrons generated by the laser field
decay into the lower-energy magnon
excitations.\cite{Munzenberg_1,Munzenberg_2,Rasing} In both cases
spin-orbit coupling (SOC) is responsible for the transfer of the
angular momentum to the lattice through different scattering
mechanism like magnon-magnon, magnon-phonon, magnon-impurity
scattering, and so on, where each process has a different
relaxation time.\cite{Arias,Hickey} The Landau-Lifshitz-Gilbert
(LLG) equation with a phenomenological damping constant $\alpha$
is commonly employed to describe magnetization dynamics of 
small-angle precessional switching.\cite{Gilbert} However, recent
studies have shown that in the case of large-angle (fast)
switching, in which the magnons are created, the LLG equation
should be extended in several aspects\cite{Brataas,Fahnle}, in
particular a $\mathbf{k}$-dependent damping constant
$\alpha_{\mathbf{k}}$ has been proposed,\cite{Vignale,ZangZang}
which allows short-wave-length magnons to relax faster than
those with $\mathbf{k}\rightarrow 0 $. Thus, the magnetization
relaxation processes, specifically the damping of magnons, play an
important role in designing ultrahigh-density magnetic recording
media.

\begin{table*}[t]
\caption{Lattice parameters $a$, spin magnetic moments
$m^{\textrm{s}}$ (in $\mu_B$), average screened on-site direct
(diagonal) ($W=\frac{1}{5}\sum_{n}^{(3d)}W_{nn;nn}$) and exchange
($J = \frac{1}{20} \sum_{m,n (m\neq n) }^{(3d)}W_{mn;nm}$) Coulomb
matrix elements (in eV) between the 3\textit{d} orbitals, and magnon
stiffness constants $D$ (in meV\AA$^2$) for tetragonal FeCo
compounds grown on Pd, Ir, and Rh. Lattice parameters are taken from
Ref.\,\onlinecite{yildiz_2}.}
\begin{ruledtabular}
\begin{tabular}{lccccccccccccc}
& $a$(\AA) & $c/a$ & $m^{\textrm{s}}_{\textrm{[Fe]}}$  &
$m^{\textrm{s}}_{\textrm{[Co]}}$  &
$m^{\textrm{s}}_{\textrm{[int]}}$  &
$m^{\textrm{s}}_{\textrm{[total]}}$  & $W_{\textrm{Fe}}$  &
$W_{\textrm{Co}}$  & $J_{\textrm{Fe}}$ & $J_{\textrm{Co}}$ &
$D_{\parallel}$ & $D_{\perp}$ & $D_{\textrm{avg}}$ \\ \hline
FeCo/Pd  & 2.75 &  1.13  & 2.81 & 1.79 & -0.12 & 4.48  & 1.68 & 1.75 & 0.52 & 0.54 & 470 & 650 & 560 \\
FeCo/Ir  & 2.72 &  1.18  & 2.80 & 1.75 & -0.13 & 4.42  & 1.68 & 1.62 & 0.52 & 0.52 & 392 & 538 & 465 \\
FeCo/Rh  & 2.69 &  1.24  & 2.80 & 1.73 & -0.13 & 4.40  & 1.68 & 1.49 & 0.52 & 0.51 & 206 & 372 & 289 \\
\end{tabular}
\label{table}
\end{ruledtabular}
\end{table*}

The aim of the present Communication is to study magnon dynamics
in tetragonal FeCo compounds from first principles. Using a
recently developed Green-function method\cite{Spex} based on the
many-body perturbation theory in the $GW$ approximation in
combination with the multiple-scattering $T$ matrix in a Wannier
basis,\cite{SW_paper,Wan_1} we have calculated the dynamical spin
susceptibility (DSS)  of tetragonal FeCo compounds whose $c/a$
ratios were fixed to the experimentally determined values that
relate to the three different substrates. As the unit cell
contains two magnetic atoms, the calculated magnon dispersions
exhibit two branches: an acoustic and an optical branch. The
former persists throughout the Brillouin zone indicating a
localized nature of magnetism in FeCo compounds. The optical
branch, on the other hand, is heavily damped due to the coupling
to single-particle Stoner excitations. We find that the tetragonal
distortion gives rise to significant magnon softening. The average
magnon stiffness constant $D$ decreases almost by a factor of two
from FeCo/Pd to FeCo/Rh, which means that acoustic magnons can be excited
much more easily in the latter material than in the former one. 
In field-driven fast magnetic switching processes, which take
place on a time scale of ns to 100 ps, the excess Zeeman energy
will be transferred to the acoustic magnons and thus magnon
stiffness constant $D$ and life time of  $\mathbf{k} \neq 0$
magnons play a decisive role determining the strength of the switching
field and switching time. The latter is limited by the damping of
magnons. Furthermore, damping also prevents "back-switching"
effect, which reduces the data rates in magnetic
recording devices.\cite{Schumacher}

To calculate the  ground-state properties of the tetragonal FeCo
compounds we use the full-potential linearized augmented
plane-wave method as implemented in the \texttt{FLEUR}
code.\cite{Fleur} The exchange-correlation potential is chosen in
the generalized gradient approximation.\cite{GGA} The muffin-tin
radii of the Fe and Co are chosen to be 2.29 a.u. A dense
$16\times16\times16$ $\mathbf{k}$-point grid is used. The
maximally localized Wannier functions are constructed with the
\texttt{Wannier90} code.\cite{Wan_1,Wannier90} The DSS is
calculated within a $T$-matrix approach\cite{SW_paper} as
implemented in the \texttt{SPEX} code\cite{Spex} using 8000
$\mathbf{k}$-points in the full Brillouin zone. We briefly review
the method here. Within many-body perturbation theory the
transverse DSS, $\chi^{-+}$, can be schematically written as
$\chi^{-+}=\chi^{-+}_{\mathrm{KS}}+\chi^{-+}_{\mathrm{KS}}T^{-+}\chi^{-+}_{\mathrm{KS}}~,$
where the first term on the right-hand side  represents the
response of the noninteracting system, i.e., the Kohn-Sham DSS.
The second term contains the $T$ matrix, which is given by $
T^{-+}=[1-W\chi^{-+}_{\mathrm{KS}}]^{-1}W\, $, where $W$ is the
screened Coulomb interaction. The $T$ matrix describes dynamical
correlation in the form of repeated scattering events of
particle-hole pairs with opposite spins and is responsible for the
formation of collective magnon excitations. Details of the
formalism, implementation, and applications to 3\textit{d}
transition metals can be found in Ref.\,\onlinecite{SW_paper}. The
DSS provides complete information on the magnetic excitation
spectrum including collective magnon modes as well as
single-particle Stoner excitations together with their respective
lifetimes.\cite{Savrasov,Ferdi,Kotani,Buczek,Samir} We note that
magnon lifetimes and Stoner excitations are not accessible within
the adiabatic approximation, a method mostly employed so far for
the calculation of the magnon dispersion within density functional
theory.\cite{Pajda}

Experimentally, FeCo alloys have been grown on the Pd, Ir, and Rh
substrates in the body-centered tetragonal structure, in which the
in-plane lattice constant is determined by the substrate and the
out-of-plane lattice constant changes so as to keep the volume
constant.\cite{yildiz_2} The experimental lattice parameters used
in the calculations are presented in Table\,\ref{table}. Yildiz
\textit{et al.} have shown that the structure remains stable for
film thicknesses of up to 15 monolayers, which is large enough to
consider it as a bulk in the context of  theoretical
modelling.\cite{Bulk} Since no experimental information is
available on the microscopic atomic order of FeCo alloys grown on
the different substrates, we assume a tetragonally distorted
CsCl-type (B2) structure derived from the known cubic bulk phase.
We note that the mechanism behind the giant uniaxial MAE observed
in tetragonal FeCo compounds has been discussed in detail in
Ref.\,\onlinecite{Burkert} and will not be dwelt on here. Indeed,
our calculated values of uniaxial MAE (results not shown) are very
similar to those reported by Burkert \textit{et
al.}\cite{Burkert}

\begin{figure}[t]
\begin{center}
\includegraphics[scale=0.49]{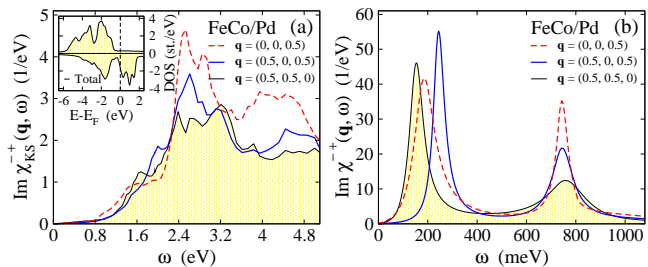}
\end{center}
\vspace*{-0.6cm} \caption{(Color online) (a) Imaginary part of the
un-renormalized Kohn-Sham spin susceptibility of the tetragonal
FeCo compounds grown on Pd ($c/a=1.13$) for selected wave vectors.
The inset shows the  spin-resolved total density of states; (b)
the same for the renormalized spin susceptibility. Note the
different scales of the axes in the two figures.
\label{fig1}}
\end{figure}

We start with a discussion of the magnetic moments and the matrix
elements of the screened Coulomb potential $W$. The latter 
are a crucial ingredient for the construction of the DSS. The
calculated values for the three different $c/a$ ratios are
presented in Table\,\ref{table}. As seen, the spin magnetic moment
of the Fe sublattice is substantially enhanced with respect to
bulk bcc (or fcc) Fe, which has a magnetic moment of about 2.2
$\mu_B$, while the Co sublattice shows moments that are more similar to
the corresponding value of bulk Co, $1.62~\mu_B$. In contrast to
bulk Co, bcc Fe is a weak ferromagnet, and thus its magnetic
moment is very sensitive to the local environment. The total spin
magnetic moment of the unit cell is around 4.4 $\mu_B$ and almost
insensitive to the tetragonal distortion, which stems from the
strong ferromagnetic nature of the FeCo compounds [see inset in
Fig.\,\ref{fig1}(a)]. Such a large magnetic moment is desirable
for magnetic recording applications as it reduces the write field
of the writing head. Not only magnetic moments but also the
average screened on-site Coulomb matrix elements $W$ (direct) and
$J$ (exchange) of FeCo compounds are insensitive to the tetragonal
distortion. The obtained values are slightly larger than the
corresponding values in the bulk phase of bcc Fe and fcc Co. As
discussed in detail in Ref.\,\onlinecite{SW_paper}, this difference
can be attributed to the larger exchange splitting of the Fe and
Co atoms in FeCo compounds as the larger the exchange splitting
the less screening takes place leading to a stronger Coulomb 
interaction $W$. Indeed, with increasing $c/a$
ratio the magnetic moment (exchange splitting) of the Co atom
decreases slightly, giving rise to a small reduction in
the Coulomb matrix elements $W_{\textrm{Co}}$ as shown in
Table\,\ref{table}.

Figure \ref{fig1}(a) presents the non-interacting Kohn-Sham
transverse DSS  of the tetragonal FeCo compounds grown on Pd
($c/a=1.13$)  for selected wave vectors at high symmetry points M
[$\mathbf{q}=(0.5,0.5,0)$], Z [$\mathbf{q}=(0,0,0.5)$], and R
[$\mathbf{q}=(0.5,0,0.5)$]. As there is no dynamical correlation
due to the  absence of electron-electron interactions, only
single-particle spin-flip Stoner excitations exist. As a
consequence, the spectral function $\textrm{Im}
\chi_{\mathrm{KS}}^{-+}(\mathbf{q},\omega)$ exhibits a broad peak
originating from  spin-flip transitions between
occupied majority and unoccupied minority states. The peak maximum 
at around 2.5 eV corresponds to the exchange splitting of the FeCo compounds
visible in the density of states shown in the inset of
Fig.\,\ref{fig1}(a). The situation is very similar for FeCo
compounds grown on Ir and Rh. As will be discussed below, Stoner
excitations are responsible for the damping of the magnons.

\begin{figure}[t]
\begin{center}
\includegraphics[scale=0.54]{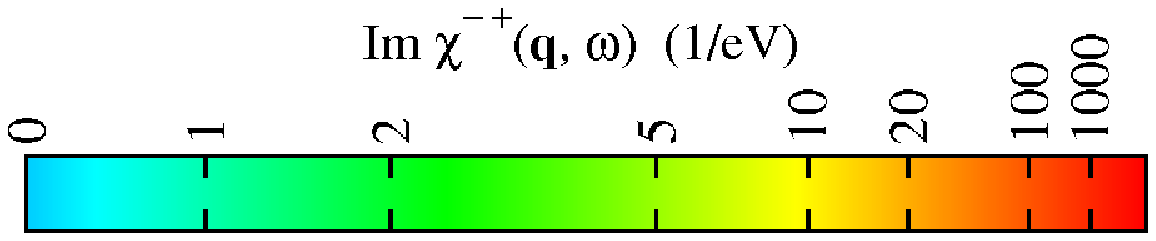}
\includegraphics[scale=0.52]{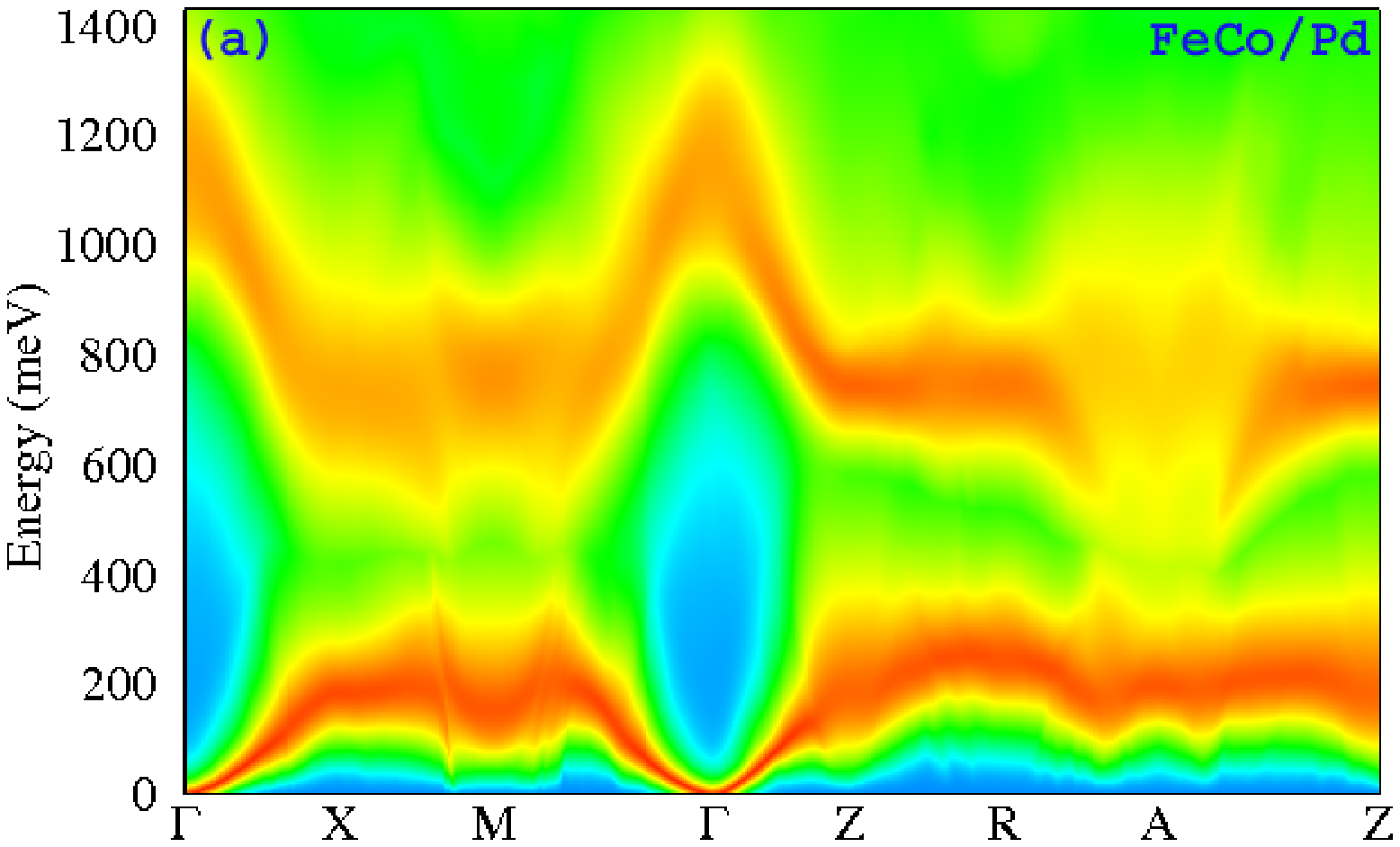}
\includegraphics[scale=0.52]{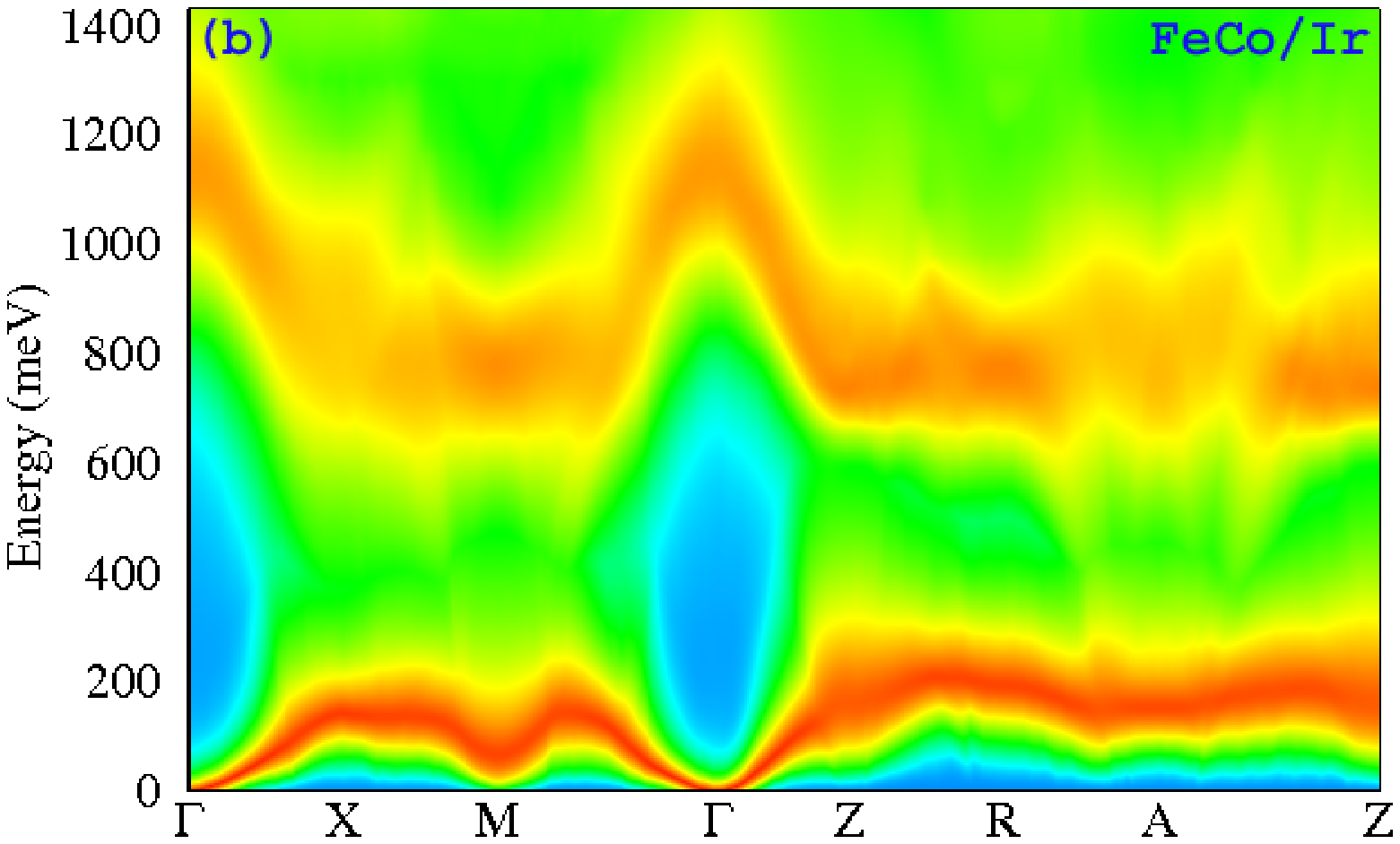}
\includegraphics[scale=0.52]{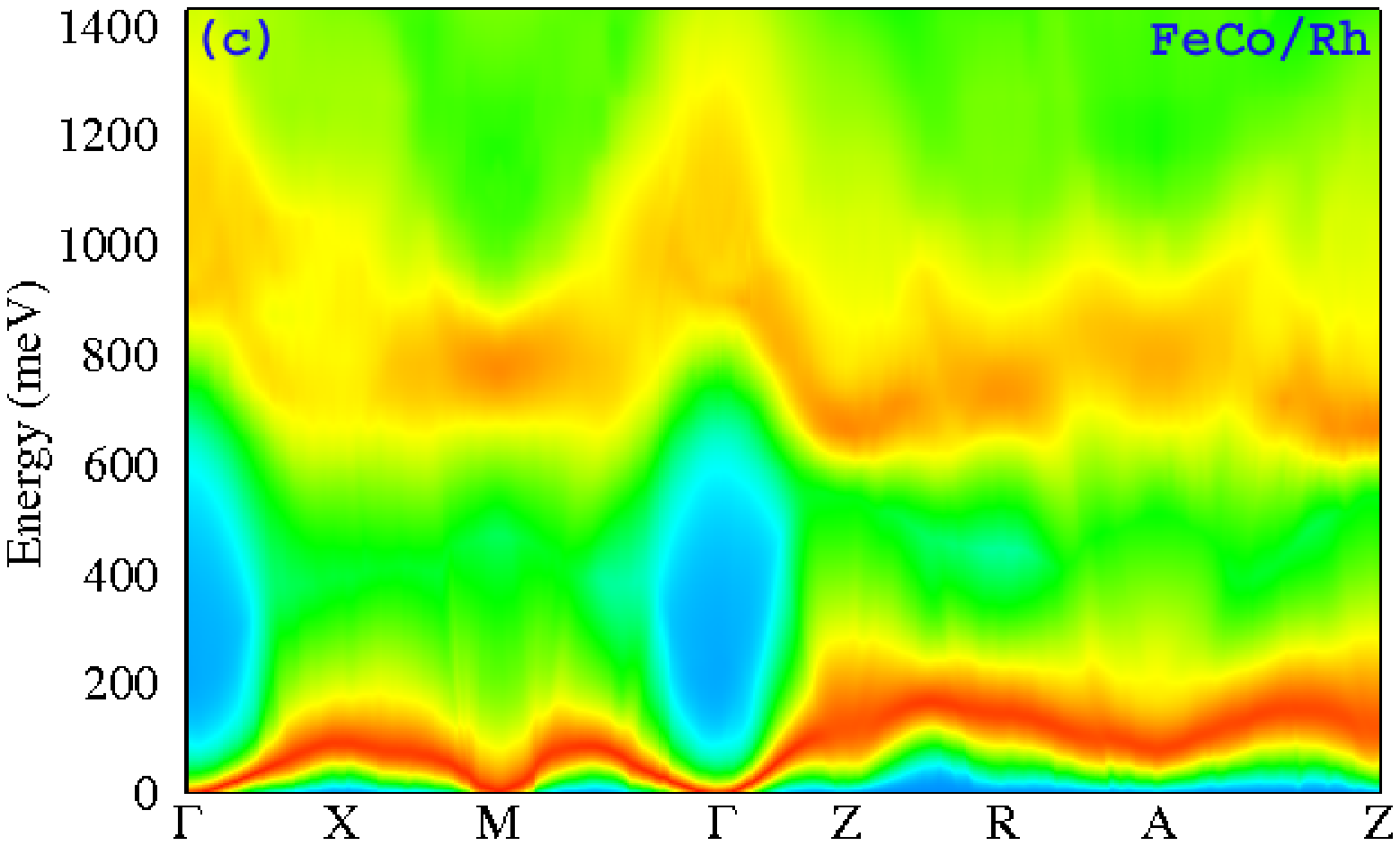}
\end{center}
\vspace*{-0.5cm} \caption{(Color online) Magnon dispersion of
the tetragonal ordered bulk  FeCo alloy as grown on (a)  Pd ($c/a=1.13$)
(b) Ir ($c/a=1.18$), (c) Rh ($c/a=1.24$), along the
high-symmetry lines in the Brillouin zone. \label{fig2}}
\end{figure}

When dynamical correlation is included in the form of interacting
spin-up and -down electrons via the screened Coulomb interaction
$W$ (see Table\,\ref{table}), additional magnon peaks appear in the
spectral function of the interacting system $\textrm{Im}
\chi^{-+}(\mathbf{q},\omega)$ at low energies as illustrated in
Fig.\,\ref{fig1}(b) for the case of FeCo/Pd. Since the unit cell
contains two magnetic atoms, we obtain two modes: a low-energy
acoustic mode and a high-energy optical mode. The former
corresponds to the in-phase precession of the Fe and Co magnetic
moments, while the latter is due to out-of-phase precession.  The
broadening of the peaks is caused by coupling to single-particle
Stoner excitations. Plotting the renormalized susceptibility as a
function of the wave vector yields the magnon dispersion as
displayed in Fig.\,\ref{fig2} along the high-symmetry lines in the
Brillouin zone for all three systems. As seen in all cases, we
obtain a well-defined low-energy acoustic branch throughout the
Brillouin zone, while the high-energy optical branch lies above
0.75 eV and is heavily damped due to coupling to Stoner
excitations. This strong damping can be traced back to the density
of states (DOS) of Stoner excitations in the respective energy
region, see Fig.\,\ref{fig1}(a). We observe a drastic increase of
the DOS above 0.8 eV. Below 0.8 eV the DOS is small because the
FeCo alloy is a strong ferromagnet, i.e., the spin-majority states
are fully occupied. Thus, though damped, the optical branch
remains identifiable in FeCo alloys, while in weak ferromagnets
like bcc Fe the magnons persist only up to 500~meV. Above this
energy they disappear due to strong coupling to Stoner
excitations.\cite{Buczek,Arno} We note in passing that
\textit{s-d} mixing leads to damping of the magnons also in cases
of strong ferromagnetism, but this effect is relatively weak
compared to damping in weak ferromagnets. Damping of magnons does
not mean that the angular momentum is transferred to the lattice. In
the absence of SOC the angular momentum stays in the magnetic subsystem,
i.e., it is transferred from magnons to the non-coherent single-particle 
spin-flip Stoner excitations. The SOC is responsible for
angular momentum transfer from the magnetic subsystem to the lattice
subsystem. We also note that we have not included  SOC in the
calculation of the DSS. The SOC gives rise to an opening of a gap in
the magnon dispersion at the $\Gamma$ point, whose value is
determined by the MAE $K_\mathrm{u}$, which is less than 1 meV for
the systems considered here.\cite{Burkert} Apart from that, we do
not expect a qualitative change in the magnon spectra because the SOC
only has a negligible effect on the electronic structure.

With increasing $c/a$ ratio the two branches are pushed in
opposite directions, i.e, the excitation energy of the optical
(acoustic) magnons increases (decreases). This behavior, on the
one hand, leads to weak damping of the acoustic  magnons since the
intensity of the Stoner DOS decreases at low energies, while the
situation is just the opposite for the optical magnons. On the
other hand, it gives rise to  magnon softening, i.e., the magnon
stiffness constant $D$ presented in Table\,\ref{table} is
considerably reduced with increasing $c/a$ ratio. As seen, from
FeCo/Pd to FeCo/Rh the average $D$ decreases almost by a factor of
two. The calculated average magnon stiffness constants $D$ for
FeCo/Pd  and FeCo/Rh are close to  the experimental  values
of fcc Co ($D=580$ meV\AA$^2$) and bcc Fe ($D=280$ meV\AA$^2$),
respectively.\cite{Pauthenet}  Furthermore, the in-plane
(D$_{\parallel}$ ) and out-of-plane (D$_{\perp}$) magnon stiffness
constants differ a lot, and this difference increases with
increasing $c/a$ ratio. The small value and anisotropy of the 
exchange stiffness constant $D$ can be attributed to a strong 
direction dependence of the exchange interactions.\cite{Ni2MnGa}
Note that in itinerant ferromagnets there are several coexisting 
exchange interactions. A detailed discussion of them for present
systems is beyond the scope of the this work. 
In the following, we will discuss them only qualitatively. In
3\textit{d} ferromagnets and their alloys the total exchange
coupling can be divided into two contributions: $
J_{\textrm{T}}=J_{\textrm{direct}} + J_{\textrm{indirect}} $,
where the first term (direct coupling) is a short-range
interaction due to the overlap of the 3\textit{d} wavefunctions
and its strength depends on the distance between the magnetic atoms,
while the long-range indirect part is due to the coupling of the
localized 3\textit{d} moments to the itinerant \textit{sp}
electrons. For $c/a \neq 1$ the $J_{\textrm{direct}}$ becomes
anisotropic. With increasing $c/a$, i.e., from FeCo/Pd to
FeCo/Rh, the in-plane and out-of-plane components of
$J_{\textrm{direct}}$ can differ a lot. The former (latter) is
expected to increase (decrease) due to smaller (larger)
interatomic distances. Consequently, the $c/a$ behavior of the
direct exchange coupling can qualitatively account for the
anisotropy of the magnetization and the reduction of the magnon
energies (acoustic branch) along the $z$ direction in FeCo
compounds. However, the strong in-plane  magnon softening is more
likely connected to $c/a$ behavior of the long range indirect
exchange interactions, which give a substantial contribution to
the total exchange coupling $J_{\textrm{T}}$ with a negative sign.
A qualitative estimate of its contribution to $J_{\textrm{T}}$ is
not easy without very detailed electronic structure analysis since
this coupling shows RKKY-type oscillations, extends over very
large distances, and is very sensitive to tetragonal
distortion.\cite{Ni2MnGa} Its strength and long-range behavior is
determined by several parameters such as conduction electron spin
polarization, Fermi surface topology, position of unoccupied
states with respect to the Fermi level and so on. For a detailed
discussion on the indirect exchange coupling in 3\emph{d}
transition metal alloys the reader is referred to
Ref.\,\onlinecite{RKKY}. Finally we would like to note that as the
magnetism in itinerant ferromagnets depends on the electronic
states far from the Fermi level, the disorder between Fe and Co
sublattices is not expected to substantially influence the magnon
spectra of FeCo compounds.\cite{Bozorth} However this is not the case for MAE,
which is very sensitive to the Fermi surface topology.\cite{Turek_CPA}

In conclusion, we have calculated the  magnon spectra of the
tetragonal bulk FeCo compounds from first principles considering
three different experimental $c/a$ ratios: FeCo grown on Pd, Ir,
and Rh with $c/a=$1.13, 1.18, and 1.24, respectively. We have
found that for all three cases the short-wave-length magnons are
strongly damped and tetragonal distortion gives rise to a
significant magnon softening. The magnon stiffness constant $D$
decreases almost by a factor of two from FeCo/Pd to FeCo/Rh, which
reduces the switching field and yields efficient excitation of the
$\mathbf{k}\neq 0$ magnons. Furthermore, the obtained strong
damping of large-wave-vector magnons in FeCo compounds suggests a
$\mathbf{k}$-dependent damping constant $\alpha_{\mathbf{k}}$ in the
LLG equation in describing magnetization dynamics of large-angle
fast precessional switching. Combination of soft magnons with
their substantial damping at large wave vectors as well as giant
MAE suggests FeCo/Rh to be a very promising material for  
ultrahigh-density perpendicular magnetic recording applications.

Fruitful discussions with  Ph. Mavropoulos, A. Schindlmayr, G.
Bihlmayer, D. B\"{u}rgler, A. Kakay, and M. C. T. D.\  M\"{u}ller
are gratefully acknowledged.

\end{document}